\documentclass[usenatbib]{mn2e}
\usepackage{natbib,psfig}

\renewcommand\[{\begin{equation}}
\renewcommand\]{\end{equation}}
\def\ex#1{\langle#1\rangle}
\def\Mpc{\,{\rm Mpc}}

\def\kms{\,{\rm km}\, {\rm s^{-1}}}

\def\d{{\rm d}}\def\e{{\rm e}}
\def\p{\partial}
\def\i{\relax\ifmmode{\rm i}\else\char16\fi}

\def\lesssim{{_ <\atop{^\sim}}}
\def\lta{\lesssim}

\def\fracj#1#2{{\textstyle{#1\over#2}}}
\def\b#1{{\bf{#1}}}

\def\lesssim{\mathrel{\hbox{\rlap{\hbox{\lower4pt\hbox{$\sim$}}}\hbox{$<$}}}}
\def\gtrsim{\mathrel{\hbox{\rlap{\hbox{\lower4pt\hbox{$\sim$}}}\hbox{$>$}}}}

\def\apj#1 #2{ApJ, #1, #2}
\def\aj#1 #2{AJ, #1, #2}
\def\mn#1 #2{MNRAS, #1, #2}
\def\aa#1 #2{A\&A, #1, #2}

\begin{document}

   \title[Rotation and anisotropy revisited]
   {Rotation and anisotropy of galaxies revisited}

   \author[J. Binney]
          {James Binney
           \\
           Rudolf Peierls Centre for Theoretical Physics, Keble Road, Oxford OX1 3NP\\
          }

%%%   \date{Received ...; accepted ...}
   \date{}

   \maketitle

\begin{abstract}
The use of the tensor virial theorem (TVT) as a diagnostic of anisotropic velocity
distributions in galaxies is revisited. The TVT provides a rigorous global
link between velocity anisotropy, rotation and shape, but the quantities
appearing in it are not easily estimated observationally. Traditionally use
has been made of a centrally averaged velocity dispersion and the peak
rotation velocity. Although this  procedure cannot be rigorously justified,
tests on model galaxies show that it works surprisingly well. With the
advent of integral-field spectroscopy it is now possible to establish a
rigorous connection between the TVT and observations. The TVT is
reformulated in terms of sky-averages, and the new formulation is tested on
model galaxies.
\end{abstract}


%=====================
\section{Introduction}
%=====================

Thirty years ago our understanding of elliptical galaxies was revolutionized by the discovery
that most giant elliptical galaxies are not flattened by rotation
\citep{BertolaC,Illingworth,Binney78}. Subsequently it emerged that a
galaxy's peak rotation speed is correlated with its luminosity, and the
cuspiness and diskiness of its luminosity density: less luminous galaxies
tend to have cuspier central density profiles, disky rather than boxy
isophotes, and a higher degree of rotational flattening than more luminous
galaxies \citep{Bender,BenderK}. 

It is generally recognized that these correlations must be clues to how
elliptical galaxies formed. A promising conjecture is that the current
configuration of a boxy galaxy was largely established by a violent merger
of galaxies of comparable mass, while cuspy galaxies were configured by a
succession of minor mergers with companions substantially less massive than
themselves \citep{NaabBH,NaabB}. To assess the truth of this and any competing
conjecture, we clearly need to characterize as fully as possible
the degree of rotational flattening of any given galaxy, and to understand
the significance of rotational flattening for the system's internal
dynamics. 

Ideally one would explore the connections between diskiness, cuspiness and
anisotropy by experimenting with a series of semi-analytic galaxy models
that would be generalizations of popular spherical models 
\citep[e.g.,][]{King,Jaffe,Hernquist}. Unfortunately, even now the repertoire of
semi-analytic flattened models is very sparse, and all of these models are
unrealistic in one significant respect or another. Moreover, thirty years
ago, when the importance of anisotropy was first realized, even fewer
semi-analytic models were available, and our theoretical understanding has
relied heavily on (a) N-body models \citep{Binney76,AarsethB,BarnesH}, (b)
Schwarzschild modelling \citep{Schwarz,Richstone}, and 
(c) the tensor virial theorem \citep{Binney78}.

The tensor virial theorem (TVT) provides a powerful general framework within
which to discuss the connection between rotation and flattening, but it is
not straightforward to connect the quantities that appear in it to
observable quantities. Until recently it was in principle impossible to make
this connection rigorously because the TVT operates at a global level, while
galaxy kinematics could be probed only along a limited number of slits, that
rarely extended very far out in the galaxy, especially perpendicular to the
apparent major axis. With the advent of integral-field spectroscopy
\citep{Bacon,SAURON,Kelz} this problem is substantially alleviated, and it is now time to
reassess how the TVT is used to interpret kinematic data. This task is the
primary goal of this paper. In the process we make a critical reassessment
of how traditional long-slit data are interpreted.

\section{The tensor virial theorem}

The TVT states that in any
equilibrium stellar system \citep{Chandrasekhar,BT}
 \[\label{TVTeq}
2\b K+\b W=0,
\]
 where the kinetic- and potential-energy tensors are
\begin{eqnarray}
K_{ij}&\equiv&\fracj12\int\d^6\b w\,v_iv_j f(\b w)\nonumber\\
W_{ij}&\equiv&-\int\d^3\b x\,\rho x_i{\p\Phi\over\p x_j}.
\end{eqnarray}
 Here $\b w\equiv(\b x,\b v)$ is the vector of phase-space coordinates, $f$
is the system's distribution function, $\rho(\b x)\equiv\int\d^3\b v\,f(\b
w)$ is the density, and $\Phi(\b x)$ is the gravitational potential. $\b K$
is customarily decomposed into contributions from ordered and random motion
 \[
\b K=\b T+\fracj12\b\Pi,
\]
 where
\begin{eqnarray}
T_{ij}&\equiv&\fracj12\int\d^3\b x\,\overline{v}_i\overline{v}_j\rho\nonumber\\
\Pi_{ij}&\equiv&\int\d^6\b w\,(v_i-\overline{v}_i)(v_j-\overline{v}_j)f(\b w).
\end{eqnarray}
 Here a bar over any quantity denotes an average over velocity space:
 \[
\overline{v}_i(\b x)\equiv{1\over\rho(\b x)}\int\d^3\b v\,v_if(\b w).
\]
 We use this notation to define the velocity-dispersion tensor
 \[\label{defssigma}
\sigma^2_{ij}\equiv\overline{(v_i-\overline{v}_i)(v_j-\overline{v}_j)}.
\]

In the case of a flattened, axisymmetric galaxy, we expect the principal
axes of the tensors above to coincide with the symmetry axis, which we label
the $z$ axis, and any two perpendicular axes -- the $x$ and $y$ axes. Then
$\b\Pi$ can be
characterized by two numbers: $\sigma_0^2$ and the global anisotropy
parameter $\delta$, which are defined such that
 \[\label{defsdelta}
\Pi_{xx}=M\sigma_0^2\quad;\quad \Pi_{zz}=(1-\delta)M\sigma_0^2,
\]
 where $M$ is the galaxy's mass.
If we assume that the galaxy's only streaming motion is rotation, we may
characterize $\b T$ by a single number $v_0$ through
 \[
T_{xx}=T_{yy}=\fracj14 Mv_0^2,
\]
 where the factor $\fracj14$ is chosen so that the total ordered kinetic
energy is $T_{xx}+T_{yy}=\fracj12Mv_0^2$. With this notation it is easy to
show that the TVT (\ref{TVTeq}) implies that \citep{Binney78}
 \[\label{TVTjjb}
{v_0^2\over\sigma_0^2}=2(1-\delta){W_{xx}\over W_{zz}}-2.
\]
 In the case that the galaxy's surfaces of constant density are all similar
spheroids of axis ratio $1-\epsilon$, the ratio $W_{xx}/W_{zz}$ on the right
of this equation is a function of $\epsilon$ and independent on the galaxy's
radial density profile \citep{Roberts}, so this equation provides a
connection between the rotation rate $v_0$, the mean-square velocity
dispersion parallel to the equatorial plane $\sigma_0^2$, and the global
anisotropy parameter $\delta$. If $v_0$ and $\sigma_0$ could be estimated
from observational measurements of the line-of-sight mean velocity and
velocity dispersion, equation (\ref{TVTjjb}) would enable one to determine
the anisotropy parameter $\delta$ from observational data.

\section{Application to traditional data}

Following \cite{Illingworth} and \cite{defis}, $\sigma_0$ has often been
identified with $\sigma_{0.5}$, the mean velocity dispersion
interior to $\sim\fracj12R_\e$, while the peak line-of-sight streaming
velocity $v_{\rm max}$ is identified with $\fracj14\pi v_0$. There is no
rigorous basis for either identification since both the velocity dispersion
and the streaming velocity are expected to depend on the location of the
line-of-sight. The factor $\fracj14\pi$ is motivated by the consideration
that if the luminosity density $j(R,z)$ scales with galactocentric distance
$(R^2+z^2)^{-3/2}$, and the streaming velocity is independent of $R$, then
when the galaxy is viewed edge-on, 
the line-of-sight streaming velocity will be a factor $\pi/4$ smaller than
the three-dimensional streaming velocity \citep{Binney78}.

\begin{figure}
\centerline{\psfig{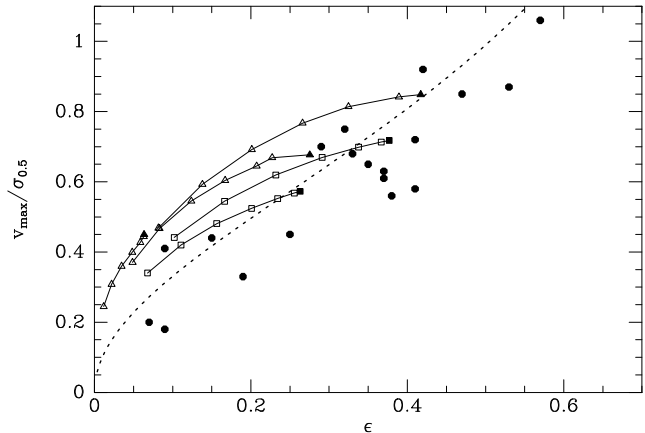}}
\caption{Squares: the ratio of peak line-of-sight streaming velocity to
line-of-sight velocity dispersion for two Evans models. Triangles: the same
data for three Rowley models. The point at the upper right of each chain
is for an edge-on model, and successive points on the chain show the
effect of reducing the inclination by $10^\circ$. The dashed curve shows
the relation (\ref{TVTjjb}) between $v_{\rm max}=\fracj14\pi v_0$ and
$\sigma_0$ for $\delta=0$. The circles show observed values of $v_{\rm
max}/\sigma_{0.5}$ for spheroids less luminous than $M_B=-20.5$.\label{TVTone}}
\end{figure} 

We can test the reliability of these traditional identifications by applying
them to models that have analytic distribution functions of the form
$f(E,L_z)$ and self-consistent gravitational potentials, where $E$ and $L_z$
are respectively the stellar energy and angular momentum about the symmetry
axis.  Fig.~\ref{TVTone} shows data for two types of semi-analytic model. 

\subsection{Evans models}

\cite{Evansmod} presented a family of models, called `power-law models',
in which the potential varies as a power of the spheroidal variable
$m=\sqrt{R_c^2+R^2+z^2/q_\Phi^2}$. Here $R_c$ is a constant that sets the
linear scale of the model, while $q_\Phi$ determines the model's flattening.
The index $y$ in the power-law relation $\Phi\propto(R_c^2/m^2)^{y}$
determines the asymptotic slope of the circular-speed curve -- the models
shown in Fig.~\ref{TVTone} both have $y=0.09$, which corresponds to a gently
falling asymptotic circular-speed curve. 

The part $f_+(E,L_z)$ of the distribution function that is even in $L_z$ is
uniquely determined by Poisson's equation, and we define the odd function
$f_-(E,L_z)$ to be equal to $f_+(E,L_z)$ for $L_z>0$. The complete
distribution function is taken to be $f=f_++\gamma f_-$. With this choice of
distribution function, the streaming velocity at any point in the model is
proportional to $\gamma$ and we choose $\gamma$ as the average of the
individual values that would make $\sigma_\phi=\sigma_R$ at a grid of points
in the equatorial plane. That is, $\gamma$ is chosen to make the model as
near as possible an isotropic rotator.

The squares in Fig.~\ref{TVTone} show centrally averaged line-of-sight
velocity dispersion and $v_{\rm max}$ for Evans models with $q_\Phi=0.85$
and $q_\Phi=0.9$. The value of $\epsilon$ is for the isophote with
semi-major axis length $15R_c$, and $\sigma_{0.5}$ is the mean value of the
line-of-sight velocity dispersion on the major axis out to $15R_c$.  The
filled squares are for a model seen edge-on, while successive open squares
show data for the models seen at inclination $i=80^\circ,70^\circ,\ldots$.

\subsection{Rowley models}

\cite{Rowley} developed  models of  flattened stellar systems to use as
model bulges. These systems are defined by their distribution function
 \[\label{Rowleyf}
f\propto\cases{\e^{\chi/\sigma^2}&for $\chi>0$,\cr0&otherwise,}
\]
 where
 \[
\chi=\chi_0-E+\omega L_z-\fracj12L_z^2/r_a^2,
\]
 with $E$ stellar energy, $L_z$ angular momentum about the symmetry axis, and
$\sigma$, $\chi_0$, $\omega$ and $r_a$ being constants.  It is easy to show
that $\chi$ can be rewritten as
 \[\label{newchi}
\chi=\Psi-\fracj12v_m^2-\fracj12\Bigl(1+{R^2\over r_a^2}\Bigr)
\Bigl(v_\phi-{\omega R\over1+R^2/r_a^2}\Bigr)^2,
\]
 where $v_m$ is the speed in the meridional plane, and
\[
\Psi\equiv\chi_0-\Phi+{\omega^2R^2\over2(1+R^2/r_a^2)} .
\]
 From equations (\ref{Rowleyf}) and (\ref{newchi}) we can identify the
streaming velocity
\[
\overline{v}_\phi={\omega R\over1+R^2/r_a^2},
\]
 and observe that the distribution of stellar velocities is a truncated
bi-axial Gaussian around $\overline{v}_\phi$. We see also that the density
vanishes for $\Psi\le0$.  We define the tidal radius $R_t$ as the radius at
which this inequality is first satisfied in the equatorial plane.
Rowley models are characterized by the dimensionless spin parameter $\omega
r_a/\sigma^2$ and the dimensionless  potential difference 
$\Delta=[\Psi(0,0)-\Psi(R_t,0)]/\sigma^2$. 

Each filled triangle in Fig.~\ref{TVTone} shows the ratio of $v_{\rm max}$
to centrally-averaged velocity dispersion versus the ellipticity of the
isophote at $3R_\e$ for a Rowley model seen edge-on, while the open
triangles show the corresponding data for the models seen at inclinations
$i=80^\circ,70^\circ,\ldots$. All models have $\Delta=4$, and $\omega
r_a/\sigma=0.9,1.35,1.69$ for the three models. 

We see that at a given ellipticity, Rowley models rotate more rapidly than
Evans models. This phenomenon reflects the low
ratio of $\sigma_\phi$ to $\sigma_R$ that follows from equation
(\ref{newchi}).  When a razor-thin disk is added to a model, the flattening
of the system is increased while its rotation stays the same, and for
realistic parameters the system can move into the band in the
($v/\sigma,\epsilon$) plane that is occupied by bulges \citep{Rowley}.

The dashed curve in Fig.~\ref{TVTone} shows the relation between
$\fracj14\pi v_0/\sigma_0$ and $\epsilon$ that is predicted by equation
(\ref{TVTjjb}).  The curve runs close to the location of edge-on Evans
models, which are very nearly isotropic rotators, while the Rowley models,
which are slightly anisotropic in the sense that
$\sigma_\phi^2<\sigma_R^2=\sigma_z^2$, lie above the dashed curve. Thus
Fig.~\ref{TVTone} validates the traditional use of the TVT to draw the
trajectory of isotropic rotators in the $(v/\sigma,\epsilon)$ plane.
The models demonstrate that tilting a galaxy away from edge-on
orientation tends to push its representative point above the
dashed curve. Hence galaxies that lie well below this curve, as
luminous elliptical galaxies nearly all do, must certainly be anisotropic in
the sense of having larger pressure parallel to the equatorial plane than
perpendicular to it.

In Fig.~\ref{TVTone} circles show ratios of peak rotation speed to velocity
dispersion averaged within $\frac12R_\e$ for real spheroidal systems less
luminous than $M_B=-20.5$ (for $H_0=50\kms\Mpc^{-1}$) from \cite{defis}.
These circles straddle the dashed curve, with rather more points below than
above it, suggesting that these systems are at most mildly anisotropic. 

\section{Application to modern data}

Observers measure the line-of-sight velocity distribution (LOSVD)
 \[
F(\b x_\perp,v_\parallel)={1\over\Sigma(\b x_\perp)}
\int\d x_\parallel\int\d^2\b v_\perp\, f(\b w),
\]
 where
\[
\Sigma\equiv\int\d x_\parallel\, \rho(\b x)
\]
 is the surface density and
the subscripts $\parallel$ and $\perp$ respectively denote components
of vectors parallel and perpendicular to the line of sight. They 
quantify $F$ by the moments
 \[
\sigma^2_\parallel(\b x_\perp)=\int\d v_\parallel\,
(v_\parallel-\widetilde{v}_\parallel)^2F(\b x_\perp,v_\parallel)
\]
 and  $\widetilde{v}_\parallel(\b x_\perp)$, 
where we have introduced the notation that for any function
$g(v_\parallel)$, 
 \begin{eqnarray}
\widetilde{g}_\parallel(\b x_\perp)&\equiv
&\int\d v_\parallel\,g(v_\parallel)
F(\b x_\perp,v_\parallel)\nonumber\\
&=&
{1\over\Sigma}\int\d x_\parallel\int\d^3\b v\, g(v_\parallel)f(\b w).
\end{eqnarray}
 Integral-field spectrographs such as SAURON \citep{SAURON} enable one
to map $\widetilde{v}_\parallel$ and $\sigma^2_\parallel$ over a
significant part of the galaxy image. From such maps it is possible to
measure quantities that are more directly related to the TVT than the
quantities $v_{\rm max}$ and $\sigma_{0.5}$ defined above.

First we note that 
\[
\sigma^2_\parallel=\widetilde{v_\parallel^2}-\widetilde{v}_\parallel^2.
\]
 Using this relation to integrate $\Sigma\widetilde{v_\parallel^2}$ over the
sky, we obtain
 \begin{eqnarray}
\int\d^2\b x_\perp\,(\sigma_\parallel^2+\widetilde{v}_\parallel^2)\Sigma
&=&\int\d^3\b x\int\d^3\b v\,v_\parallel^2f(\b w)\nonumber\\
&=&2\hat\b s\cdot\b K\cdot\hat\b s,
\end{eqnarray}
 where $\hat\b s $ is the unit vector parallel to the line of sight.

 For $\hat\b s$ parallel to the $x$ axis we can write this equation more
compactly as
 \[\label{givesKxx}
\left(\ex{\sigma_\parallel^2}+\ex{\widetilde{v}_\parallel^2}\right)M=
2K_{xx},
\] 
 where we have introduced the notation that with $q(\b x_\perp)$ an
arbitrary function on the sky, its sky-average is
 \[
\ex{q}\equiv{1\over M}\int\d^2\b x_\perp\,q\Sigma.
\]
 Equation (\ref{givesKxx}) for $K_{xx}$ enables us to derive from
the $xx$ and $zz$ components of the TVT
\[\label{TVTnew}
M{\ex{\sigma_\parallel^2}+\ex{\widetilde{v}_\parallel^2}\over\Pi_{zz}}
={W_{xx}\over W_{zz}}.
\] 

At any point $\b x$ in the galaxy let $u(\b x)$ be the difference between the component
$\overline{v}_\parallel$ of the streaming velocity parallel to the line of
sight and the mean velocity for that line of sight,
$\widetilde{v}_\parallel$. With this definition,
$u\equiv\overline{v}_\parallel-\widetilde{v}_\parallel$, it is easy to show that
 \[
\Sigma\sigma^2_\parallel=\int\d x_\parallel\,(\sigma_{xx}^2+u^2)\rho(\b x),
\]
 where $\sigma_{xx}$ is defined by equation (\ref{defssigma}).
Integrating this expression  over the sky, we obtain
 \[
\Pi_{xx}=M\ex{\sigma_\parallel^2}-\int\d^3\b x\,u^2(\b
x)\rho(\b x).
\]
 Finally we use this equation and equation (\ref{defsdelta}) to eliminate
$\Pi_{zz}$ from equation (\ref{TVTnew}), and  obtain after some rearrangement
 \[\label{modTVT}
{\ex{\widetilde{v}_\parallel^2}\over\ex{\sigma_\parallel^2}}=
{(1-\delta)W_{xx}/W_{zz}-1\over\alpha(1-\delta)W_{xx}/W_{zz}+1},
\]
 where
\[\label{defsalpha}
\alpha\equiv{1\over M\ex{\widetilde{v}_\parallel^2}}\int\d^3\b x\,u^2\rho(\b x).
\]
 In this form of the TVT, the left side contains only observationally
accessible quantities, while the right side contains the
standard global anisotropy parameter $\delta$, the shape parameter
$W_{xx}/W_{zz}$, and a new dimensionless parameter $\alpha$, which
quantifies the contribution of streaming motion to the line-of-sight
velocity dispersion.

\begin{figure}
\psfig{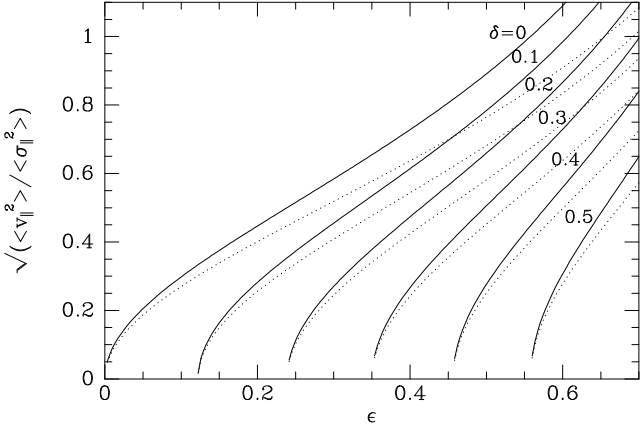}
\caption{Modified TVT. The full curves show the relation between
$\ex{\widetilde{v}_\parallel^2}/\ex{\sigma_\parallel^2}$ established by
equation (\ref{modTVT}) when  $\alpha=0$, while the dotted curves show the
relation when $\alpha=0.2$.
\label{TVTtwo}}
\end{figure}

The value of $\alpha$ depends on the shape of the galaxy's intrinsic
rotation curve, and its radial luminosity profile, but not on the amplitude
of the rotation curve. One may straightforwardly show that it vanishes in
the case of solid-body rotation, and increases with the shear of the stellar
flow. For a fixed rotation curve, it is larger for flatter slopes of the
radial luminosity density profile. A numerical calculation for the case of a
galaxy in which the luminosity density follows the \cite{Hernquist} profile
shows that $\alpha=0.131$ if $\overline{v}_\phi=\hbox{constant}$.

Fig.~\ref{TVTtwo} shows $v/\sigma$ as a function of $\epsilon$ from equation
(\ref{modTVT}) for several values of $\delta$ and two values of $\alpha$:
$\alpha=0$ (full curves) and $\alpha=0.2$ (dotted curves).  When $\alpha=0$
the right side of equation (\ref{modTVT}) becomes precisely half of the
right side of the classical result (\ref{TVTjjb}). The factor half arises
because in equation (\ref{TVTjjb}) $v_0$ is a three-dimensional streaming
velocity, whereas in equation (\ref{modTVT})
$\sqrt{\ex{\widetilde{v}_\parallel^2}}$ is a projected velocity.

Probably equation (\ref{modTVT}) is most usefully recast as an expression
for the anisotropy parameter $\delta$ as a function of $\alpha$, $\epsilon$ and 
$\ex{\widetilde{v}_\parallel^2}/\ex{\sigma_\parallel^2}$:
 \[\label{modTVTb}
\delta=1-{1+\ex{\widetilde{v}_\parallel^2}/\ex{\sigma_\parallel^2}
\over
1-\alpha\ex{\widetilde{v}_\parallel^2}/\ex{\sigma_\parallel^2}}
\left({W_{zz}\over W_{xx}}\right).
\]

\subsection{Practical considerations}

Three issues arise when using observational data to evaluate the right side
of equation (\ref{modTVTb}).  We cannot directly measure $\alpha$, but we
can infer its value with reasonable accuracy from the shape of the measured
rotation curve. In the tests below, I simply use the values directly
calculated from the models. The second issue is that our kinematic data do
not extend to arbitrary radii, so it is not practicable to evaluate the
means $\ex{\widetilde{v}_\parallel^2}$ and $\ex{\sigma_\parallel^2}$. In
practice our averages must be confined to some inner region and we must
investigate the seriousness of the error incurred by this confinement.  The
final problem is how to determine the required ratio $W_{zz}/W_{xx}$ of
components of the potential-energy tensor.  In principle this may be done
from the photometry \citep[e.g.][]{BinneyS}.  A much simpler alternative is
to assume that the isodensity surfaces have a constant ellipticity
$\epsilon$, which allows $W_{zz}/W_{xx}$ to be determined independently of
the radial density profile. In most of the tests below I have used this
crude approach, with $\epsilon$ taken to be a weighted average
$\overline{\epsilon}$ of the ellipticities of individual isophotes. After
some experimentation, the weighting scheme adopted was
 \[
\overline{\epsilon}={\int\d R\,R^2\Sigma(R,0)\epsilon(R)\over
\int\d R\,R^2\Sigma(R,0)},
\]
 where $\Sigma(R,0)$ denotes the surface brightness distance $R$ down the
 apparent major axis.

Insight into the uncertainties that will
be encountered when using real data is provided by applying equation
(\ref{modTVTb}) to pseudo-data obtained by projecting Evans and Rowley
models.

\begin{figure}
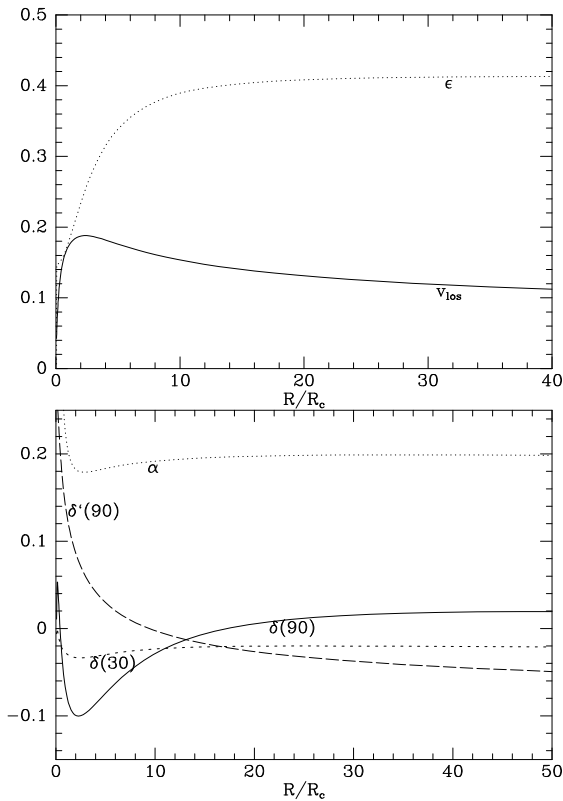

\centerline{\psfig{file=evansrot.ps,width=.9\hsize}}
\centerline{\psfig{file=plotevans.ps,width=.9\hsize}}
\caption{Upper panel: projected rotation speed and ellipticity for an Evans
model. Lower panel: the parameter $\alpha$ defined by equation
(\ref{defsalpha}) (dotted curve) and the global anisotropy parameter
$\delta$ estimated from equation (\ref{modTVTb}) using only data from within
the circle of radius $R$; at edge-on orientation (full curve) and for
inclination $i=30^\circ$ (short-dashed curve). The long-dashed curve shows
values of $\delta$ recovered at edge-on orientation when the true ratio
$W_{zz}/W_{xx}$ is used in (\ref{modTVTb}) instead of an estimate based on
the assumption that all isodensity surfaces are similar spheroids.
\label{Evansfigb}}
\end{figure}

\begin{figure}
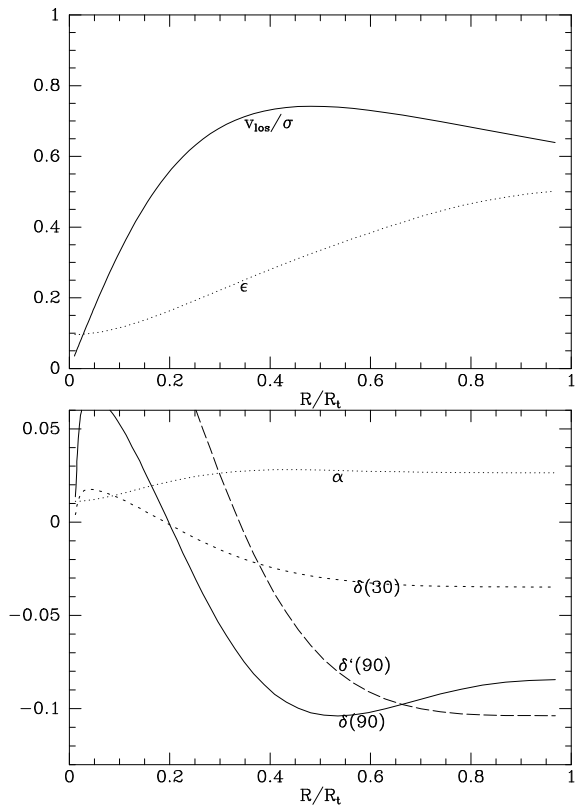

\centerline{\psfig{file=rowleyrot.ps,width=.9\hsize}}
\centerline{\psfig{file=plotrowley.ps,width=.9\hsize}}
\caption{As Fig.~\ref{Evansfigb} but for a Rowley model.\label{Rowleyfigb}}
\end{figure}

The upper panel of Fig.~\ref{Evansfigb} shows the projected rotation rate
and ellipticity of an Evans model with $y=0.25$ and $q_\Phi=0.9$.\footnote{The
final integral in the definition of $W_{ij}$ diverges for Evans models with
$y<0.25$.} Because the rotation curve rises very steeply near
the centre (in three dimensions $\overline{v}$ does not tend to zero as the
origin is approached), $\alpha\approx0.2$ is relatively large: the dotted curve
in the lower panel of Fig.~\ref{Evansfigb} shows as a function of radius $R$
the value of $\alpha$ obtained by carrying the integrals involved in its
definition (\ref{defsalpha}) only out to $R$. The full curve in the lower
panel shows as a function of $R$ the values of $\delta$ yielded by equation
(\ref{modTVTb}) when the model is seen edge-on and $W_{zz}/W_{xx}$ is
evaluated under the assumption of constant ellipticity, $\overline{\epsilon}$.  Small positive
values of $\delta<0.075$ are obtained, and outside the core (where $\alpha$
becomes large) $\delta$ rises slowly with $R$ because $\epsilon$ is rising,
while the streaming velocity is very nearly constant. The long-dashed curve
shows the values of $\delta$ obtained when $W_{zz}/W_{xx}$ is set equal to
the value it assumes at large radii. The decline of this curve reflects the
outward-increasing nature of $\epsilon$. The fact that at $R\simeq50R_c$
this curve has fallen to small negative $\delta\simeq-0.05$ suggests that
the procedure intended to make the model an isotropic rotator was not wholly
successful. The offset between the end points of the full and long-dashed
curves gives an indication of the error in $\delta$ that is inherent in
adopting a constant ellipticity.

Fig.~\ref{Rowleyfigb} shows equivalent data for the Rowley model that has
$\Delta=4$ and $\omega r_a/\sigma=1.69$. Because this model has a
substantial region of near solid-body rotation, $\alpha\lta 0.03$ is much
smaller than in the case of the Evans model. When the model is viewed
edge-on, $\delta$ falls by $0.35R_t$ to $\delta=-0.075$ and from there to
the edge of the model oscillates in a small range around $-0.097$. The
end-point of the long-dashed curve, which is based on the exact ratio
$W_{zz}/W_{xx}$, lies within this range because at most radii the azimuthal
velocity dispersion is smaller than the other two principal dispersions.

The short-dashed curves in the lower panels of Figs~\ref{Evansfigb} and
\ref{Rowleyfigb} show that remarkably small values of $\delta$ are obtained
when equation (\ref{modTVTb}) is applied to data obtained from the models at
inclination $i=30^\circ$. Since equation (\ref{modTVTb}) was derived for
edge-on inclination, we have no guarantee that a realistic value of $\delta$
will be obtained at small inclinations. Moreover, in Fig.~\ref{TVTone} both
models move away from the isotropic-rotator line towards negative $\delta$
as the inclination is reduced. Hence it comes as a pleasant surprise to find
that equation (\ref{modTVTb}) works well at small inclinations.

\section{Conclusions}

The degree of anisotropy in the velocity distribution of an elliptical
galaxy is probably an important clue to the manner in which the galaxy
formed. Traditionally anisotropy has been estimated from the ratio $v_{\rm
max}/\sigma_{0.5}$ of the maximum measured rotation velocity to a central
mean of the velocity dispersion. By a distinctly ad-hoc scaling, the tensor
virial theorem has been used to draw a curve on the $(v/\sigma,\epsilon)$
plane for isotropic rotators, and galaxies that lie significantly below this
curve have been deemed to have anisotropic velocity distributions.
Fig.~\ref{TVTone} validates this procedure by showing the models with nearly
isotropic velocity distributions are placed near the curve of isotropic
rotators.

A much more rigorous procedure is possible now that one can map rotation
speed and velocity dispersion over a substantial fraction of a galaxy's
image. One replaces $v_{\rm max}/\sigma_{0.5}$ with the ratio
$\ex{\widetilde{v}_\parallel^2}/\ex{\sigma_\parallel^2}$ of the sky-averaged
squared rotation velocity to the squared velocity dispersion. Equation
(\ref{modTVT}) rigorously relates this average to the global anisotropy
parameter $\delta$, a shape-dependent ratio of potential-energy tensor
components, and the dimensionless quantity $\alpha$ that measures the degree
of shear in the stellar streaming velocity.  Alternatively, equation
(\ref{modTVTb}) expresses the anisotropy parameter as a function of
quantities that are either directly observable
($\ex{\widetilde{v}_\parallel^2}/\ex{\sigma_\parallel^2}$ and
$W_{zz}/W_{xx}$) or can be estimated to sufficient accuracy from the data
($\alpha$).

An obvious problem with the rigorous approach is that the sky-averages that
appear in the TVT extend over the whole image, while data are available only
within some limiting radius $R$. Figs~\ref{Evansfigb} and \ref{Rowleyfigb}
suggest that $\delta$ can be estimated to an uncertainty $\simeq0.05$ from
data that extend out to $\ga15R_c$ in a galaxy that resembles an Evans model
(in which the ellipticity reaches a plateau around $10R_c$), or $\ga0.35
R_t$ in a galaxy like a Rowley model (in which $\epsilon $ roughly linearly
to the edge of the system). It is unfortunate that when we confine ourselves
to semi-analytic models, we are able to investigate the impact of restricted
sky coverage only with models that have $\delta\simeq0$.  Tests
on systems with $\delta\gg0$ could be carried out with N-body models.

Traditionally, anisotropy has been quantified as the ratio $(v/\sigma)^*$ of
the measured quantity $v_{\rm max}/\sigma_{0.5}$ to the height of the
isotropic-rotator curve at a value of $\epsilon$ that is characteristic of
the galaxy. This work suggests that anisotropy can in future be more
rigorously quantified by the anisotropy parameter $\delta$.

\end{document}